# Decode and Forward Relaying for SC-FDE Systems with a Multi-Antenna Relay

Farshad Dizani, Ali Olfat
Signal Processing and Communication Systems Laboratory, University of Tehran, Tehran, Iran
f.dizani@ut.ac.ir, aolfat@ut.ac.ir

*Abstract*—In this paper, a cooperative relay network consisting of a single-antenna source, a multi-antenna relay, and a multi-antenna destination is considered. The relay operates in decode-and-forward (DF) mode under frequency-selective fading. To combat intersymbol interference (ISI), single-carrier frequency-domain equalization (SC-FDE) with or without decision feedback is deployed at the relay and the destination. The equalization coefficients are obtained using minimum mean squared error (MMSE) criterion. Both equal and optimum power allocations for a constant total transmit power at the relay are considered. While, the optimum power allocation is a non-convex problem, the solution is obtained using strong duality.

## I. INTRODUCTION

Cooperative transmission for combating fading through spatial diversity has attracted enormous attention recently [1]-[2]. Two major and popular protocols in relaying are amplify-and-forward (AF) and decode-and-forward (DF) [3]. In AF, the nodes amplify their received signal and retransmit them to the destination. Whereas in DF, the nodes decode their received signal and retransmit them.

Future wireless communication systems need high data rate transmission which translates into large channel impulse responses (CIRs) in practice. In frequency-selective channels, time-domain equalization suffers from high complexity [4]. We can overcome this problem using frequency-domain equalization methods. Orthogonal frequency-division multiplexing (OFDM) and single-carrier frequency-domain-equalization (SC-FDE) are well known for this goal. Some disadvantages of OFDM systems are high peak-to-average power ratio (PAPR) and sensitivity to carrier frequency offset (CFO) which motivate to the employment of SC-FDE transmission systems [5]. SC-FDE has been adopted for uplink transmission and OFDM for the downlink in LTE-advanced [6].

In cooperative communication, most of the proposed methods in literature are considered in flat fading. For frequency-selective channels major efforts have been focused on OFDM transmission and the literature on cooperative transmission for SC-FDE is very poor, especially in DF protocol [7]-[8]. In [7], the asymptotic outage probability with the aim of achieving the maximum diversity order is derived in DF relaying SC-FDE system with multiple sources, multiple relays with best relay selection, and a single destination. The authors of [8] have proposed a single-carrier frequency-domain equalizer and diversity combining method with DF relaying over frequency-selective channels. They have employed multiple relays with linear FDE and applied diversity combining at the destination. In their relaying scheme, full channel state information (CSI) is required at the destination, nevertheless, we will show that our relaying scheme has better performance without needing full CSI.

In this paper, we employ SC-FDE in DF cooperative relay networks under frequency-selective fading. Considering one single-antenna source, one multi-antenna relay, and a multi-antenna destination, the optimum filter coefficients at the relay and destination are derived using minimum mean squared error (MMSE) criterion. The coefficients at the destination are obtained under two strategies for power allocation, i.e., equal and optimum power allocations for a constant total transmit power at the relay. In most of proposed AF relaying methods for SC-FDE transmission, full CSI is needed, whereas full CSI is not required in our DF relaying scheme.

The rest of the paper is organized as follows. In Section II, the system model is described. Optimum coefficients for relay filters are obtained using MMSE criterion in Section III. In Section IV, the optimum coefficients for filters at the destination are obtained. Simulation results are provided in Section V and finally, we have drawn our conclusions in Section VI.

*Notation*: In this paper, $A^{-1}, A^*, A^T$, and $A^H$ denote the inverse, conjugate, transpose, and conjugate transpose of matrix $A$, respectively. Moreover, $E\{\cdot\}, |\cdot|, \nabla$, and $\mathbf{I}_{N \times N}$ denote statistical expectation, absolute value, gradient, and $N \times N$ identity matrix.

## II. SYSTEM MODEL

We consider one-way relay networks with one source node (S), one multi-antenna relay (R), and a multi-antenna destination (D). In our system model, there is no direct link between the source and the destination. A block diagram of the overall transmission is shown in Fig. 1.

According to Fig. 1, $N_R$ and $N_D$ are the number of antennas at the relay and the destination, respectively. We employ DF relaying in which the transmission is organized in two phases. In the first phase, the source node transmits its data to the relay and in the second phase, the relay forwards the decoded data to the destination node.

The channels are supposed frequency selective in both the source-relay and relay-destination links. We define $h_i[k], 0 \le k \le L_h - 1$ and $g_{i,j}[k], 0 \le k \le L_g - 1$ as discrete-time CIRs between the source and the $i$th antenna of relay and between the $i$th antenna of relay and the $j$th antenna of destination, respectively. $L_h$ and $L_g$ denote the length of the source-relay and the relay-destination channels, respectively. To combat intersymbol interference (ISI), we use SC-FDE at both the relay and the destination. In what follows, the two phases are described in detail.

### A. First Phase

Source node transmits its information to the relay during the first phase. In each transmission, $M$ independent and identically distributed (i.i.d.) symbols with variance $\sigma_s^2 = P_S$ are used. Before transmission, a cyclic prefix (CP) is added at the beginning of the $M$ symbols. Cyclic prefix is the last $L_{CP}$ symbols ($L_{CP} \geq L_h - 1$) that prevents interference with the previous transmission and also creates the circular convolution of the CIRs and transmitted signals. After transmission, the first $L_{CP}$ received symbols should be discarded at the relay. The received signal at the $i$th antenna of relay is

$$r_{m,i} = \sum_{k=0}^{L_h-1} s_{m-k} h_i[k] + n_{m,i}, \quad (1)$$

for $m = 0, 1, \ldots, (M-1)$, where $n_{m,i}$ denotes additive noise at the $i$th antenna of relay and $s_m = s_{M+m}$ for $m = -L_{CP}, -L_{CP}+1, \ldots, -1$, due to the presence of the CP. Additive noise is assumed to be i.i.d. zero-mean complex Gaussian random variable with variance $\sigma_n^2$. In the discrete frequency-domain, (1) for $l = 0, 1, \ldots, (M-1)$ becomes

$$R_{l,i} = H_{l,i} S_l + V_{l,i}, \quad (2)$$

where

$$H_{l,i} = \sum_{m=0}^{L_h-1} h_i[m] \exp\left(-j2\pi \frac{ml}{M}\right), \quad (3a)$$

$$S_l = \sum_{m=0}^{M-1} s_m \exp\left(-j2\pi \frac{ml}{M}\right), \quad (3b)$$

and

$$V_{l,i} = \sum_{m=0}^{M-1} n_{m,i} \exp\left(-j2\pi \frac{ml}{M}\right). \quad (3c)$$

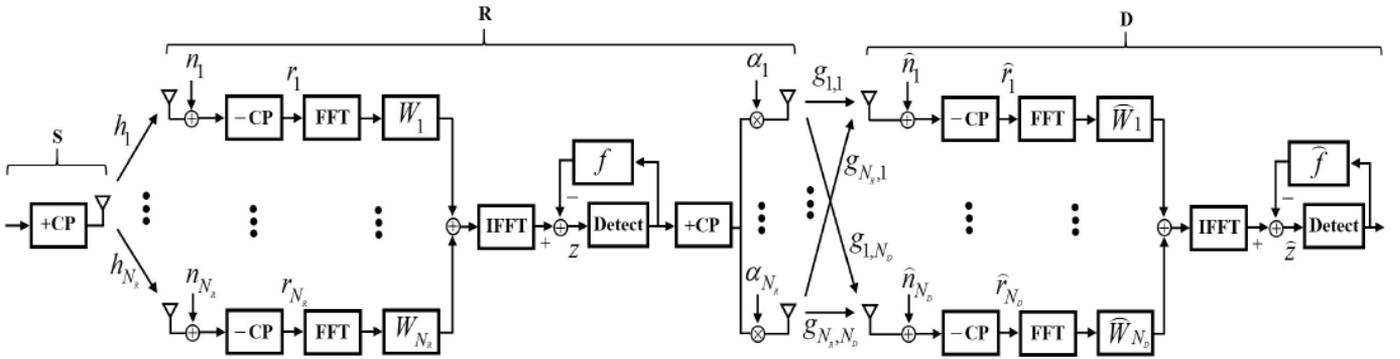

Fig. 1. System model

We consider two schemes for decoding the symbols in the first phase of transmission:

*1) FDE without time-domain decision feedback equalization (DFE):* In this scheme, by employing FDE, the output for $m = 0, 1, \ldots, (M-1)$ is

$$z_m = \frac{1}{M} \sum_{l=0}^{M-1} \sum_{i=1}^{N_R} R_{l,i} W_{l,i} \exp\left(j2\pi \frac{ml}{M}\right), \quad (4)$$

where $W_{l,i}$ for $l = 0, 1, \ldots, (M-1)$ are frequency-domain filter coefficients at the $i$th antenna of relay. According to Fig. 1, $M$-point fast Fourier transform (FFT) and inverse fast Fourier transform (IFFT) are used for FDE.

*2) FDE with time-domain DFE:* The output after using FDE with time-domain DFE for $m = 0, 1, \ldots, (M-1)$ is

$$z_m = \frac{1}{M} \sum_{l=0}^{M-1} \sum_{i=1}^{N_R} R_{l,i} W_{l,i} \exp\left(j2\pi \frac{ml}{M}\right) - \sum_{k \in F_{B_h}} f_k^* \widehat{s}_{m-k}, \quad (5)$$

where $\widehat{s}_m$ is the decoded data, $f_k^*$ is the $k$th feedback filter coefficient, and $F_{B_h}$ is a set of non-zero indices that correspond to the delays of the $B_h$ feedback coefficients ($B_h \leq L_h - 1$). In both schemes, the output is decoded and the decoded data is used in the second phase of transmission.

*B. Second Phase*

In the second phase, the relay transmits the decoded symbols in the previous phase to the destination. CP is added at the beginning of $M$ symbols and will be discarded after transmission at the destination ($L_{CP} \geq L_g - 1$). The received signal at the $j$th antenna of destination is

$$\widehat{r}_{m,j} = \sum_{i=1}^{N_R} \sum_{k=0}^{L_g-1} \alpha_i \widehat{s}_{m-k} g_{i,j}[k] + \widehat{n}_{m,j}, \quad (6)$$

for $m = 0, 1, \ldots, (M-1)$, where $\alpha_i$ is a coefficient assigned to the $i$th antenna of relay for power allocation. The discrete frequency-domain version of (6) for $l = 0, 1, \ldots, (M-1)$ is

$$\widehat{R}_{l,j} = \sum_{i=1}^{N_R} \alpha_i G_{l,i,j} \widehat{S}_l + \widehat{V}_{l,j}. \quad (7)$$

We consider two schemes for equalization and two strategies for power allocation.

*1) FDE (without DFE):* By employing FDE, the output for $m = 0, 1, \ldots, (M-1)$ is

$$\widehat{z}_m = \frac{1}{M} \sum_{l=0}^{M-1} \sum_{j=1}^{N_D} \widehat{R}_{l,j} \widehat{W}_{l,j} \exp\left(j2\pi \frac{ml}{M}\right), \quad (8)$$

where $\widehat{W}_{l,j}$ for $l = 0, 1, \ldots, (M-1)$ are frequency-domain filter coefficients at the $j$th antenna of destination.

*2) FDE-DFE:* The output after using FDE with time-domain DFE for $m = 0, 1, \ldots, (M-1)$ is

$$\widehat{z}_m = \frac{1}{M} \sum_{l=0}^{M-1} \sum_{j=1}^{N_D} \widehat{R}_{l,j} \widehat{W}_{l,j} \exp\left(j2\pi \frac{ml}{M}\right) - \sum_{k \in F_{B_g}} \widehat{f}_k^* \widehat{\widehat{s}}_{m-k}, \quad (9)$$

where the feedback filter has $B_g$ coefficients ($B_g \leq L_g - 1$). The result of the second phase of transmission is the system output. Optimum and equal power allocations are used in these schemes. These assumptions will be discussed in detail in Section IV.

### C. Required CSI and Feedback

We assume that the relay estimates the source-relay CIRs $h_i[k]$, $0 \leq k \leq L_h - 1$, $1 \leq i \leq N_R$ and the destination estimates the relay-destination CIRs $g_{i,j}[k]$, $0 \leq k \leq L_g - 1$, $1 \leq i \leq N_R$, $1 \leq j \leq N_D$, where the channel estimations are assumed perfect. In equal power allocation at the relay, the above assumption is sufficient for deriving the optimum relay filter coefficients. But in optimum power allocation, destination computes the assigned coefficients for power allocation and sends them back to the relay. Hence, full CSI is not required at the relay or destination.

## III. Decoding at the Relay

For finding the optimum frequency-domain filter coefficients, we adopt MMSE criterion in both FDE and FDE-DFE schemes.

### A. FDE

According to (2) and (4), the MMSE criterion for finding the FDE coefficients at the relay is expressed as

$$\min_{W_{l,i}} E\{|z_m - s_m|^2\} = \min_{W_{l,i}} E\left\{\left|\frac{1}{M}\sum_{l=0}^{M-1}\sum_{i=1}^{N_R}(H_{l,i}S_l + V_{l,i})W_{l,i}\exp\left(j2\pi\frac{ml}{M}\right) - s_m\right|^2\right\}. \quad (10)$$

The optimum FDE coefficients can be obtained by setting the derivative of (10) with respect to $W_{l,i}^*$ equal to zero for $l = 0, 1, \ldots, (M-1)$ and $i = 1, 2, \ldots, N_R$ as

$$W_{l,i} = \frac{H_{l,i}^* - \sum_{\substack{k=1 \\ k \neq i}}^{N_R} W_{l,k} H_{l,i}^* H_{l,k}}{\frac{\sigma_n^2}{P_S} + |H_{l,i}|^2}. \quad (11)$$

Define

$$\text{SNR} \triangleq \frac{P_S}{\sigma_n^2}, \quad (12a)$$

$$\mathbf{W}_l \triangleq (W_{l,1}, W_{l,2}, \ldots, W_{l,N_R})^T, \quad (12b)$$

and

$$\widetilde{\mathbf{H}}_l \triangleq (H_{l,1}, H_{l,2}, \ldots, H_{l,N_R})^H, \quad (12c)$$

where $l = 0, 1, \ldots, (M-1)$. Rewriting (11) for $l = 0, 1, \ldots, (M-1)$ using (12), we have

$$\mathbf{W}_l = \left(\frac{1}{\text{SNR}}\mathbf{I}_{N_R \times N_R} + \widetilde{\mathbf{H}}_l \widetilde{\mathbf{H}}_l^H\right)^{-1} \widetilde{\mathbf{H}}_l = \frac{\widetilde{\mathbf{H}}_l}{\frac{1}{\text{SNR}} + \widetilde{\mathbf{H}}_l^H \widetilde{\mathbf{H}}_l}. \quad (13)$$

### B. FDE-DFE

In this scheme, the MMSE criterion can be expressed as

$$\min_{W_{l,i}} E\left\{\left|\frac{1}{M}\sum_{l=0}^{M-1}\sum_{i=1}^{N_R}(H_{l,i}S_l + V_{l,i})W_{l,i}\exp\left(j2\pi\frac{ml}{M}\right) - \sum_{k \in F_{B_h}} f_k^* s_{m-k} - s_m\right|^2\right\}. \quad (14)$$

The optimum FDE and DFE coefficients can be obtained by setting the derivative of (14) with respect to $W_{l,i}^*$ and $f_k^*$ equal to zero as

$$W_{l,i} = \frac{H_{l,i}^*\left(1 + \sum_{k \in F_{B_h}} f_k^* \exp\left(-j2\pi\frac{kl}{M}\right)\right) - \sum_{\substack{k=1 \\ k \neq i}}^{N_R} W_{l,k} H_{l,i}^* H_{l,k}}{\frac{1}{\text{SNR}} + |H_{l,i}|^2} \quad (15)$$

and

$$f_k = \frac{1}{M}\sum_{l=0}^{M-1}\sum_{i=1}^{N_R} W_{l,i}^* H_{l,i}^* \exp\left(-j2\pi\frac{kl}{M}\right), \quad (16)$$

where $l = 0, 1, \ldots, (M-1)$, $i = 1, 2, \ldots, N_R$, and $k = k_1, k_2, \ldots, k_{B_h}$. By some calculations, we obtain

$$\mathbf{W}_l = \frac{\widetilde{\mathbf{H}}_l}{\frac{1}{\text{SNR}} + \widetilde{\mathbf{H}}_l^H \widetilde{\mathbf{H}}_l}\left(1 + \sum_{k \in F_{B_h}} f_k^* \exp\left(-j2\pi\frac{kl}{M}\right)\right), \quad (17)$$

for $l = 0, 1, \ldots, (M-1)$ and

$$\mathbf{f} = -\mathbf{V}^{-1}\mathbf{v}, \quad (18a)$$

where

$$\mathbf{f} = \left(f_{k_1}, f_{k_2}, \ldots, f_{k_{B_h}}\right)^T, \quad (18b)$$

$$\mathbf{v} = \left(v_{k_1}, v_{k_2}, \ldots, v_{k_{B_h}}\right)^T, \quad (18c)$$

$$\mathbf{V} = \begin{bmatrix} v_0 & v_{k_1-k_2} & \cdots & v_{k_1-k_{B_h}} \\ v_{k_2-k_1} & v_0 & \cdots & v_{k_2-k_{B_h}} \\ \vdots & \vdots & \ddots & \vdots \\ v_{k_{B_h}-k_1} & v_{k_{B_h}-k_2} & \cdots & v_0 \end{bmatrix}, \quad (18d)$$

and

$$v_k = \frac{1}{M \times \text{SNR}} \sum_{l=0}^{M-1} \frac{\exp\left(-j2\pi\frac{kl}{M}\right)}{\frac{1}{\text{SNR}} + \widetilde{\mathbf{H}}_l^H \widetilde{\mathbf{H}}_l}. \quad (18e)$$

## IV. Power Allocation in the Relay and Decoding at the Destination

As was mentioned before, we have two schemes for equalization at the destination and two strategies for power allocation. First, the equalization schemes will be expressed for equal power allocation and then, will be discussed for optimum power allocation.

### A. Equal Power Allocation

Suppose that the total power at the relay is $P_R = \widetilde{\sigma}_s^2$ and the additive noise variances at the antennas of destination are all $\widehat{\sigma}_n^2$. For dividing the relay power equally between its antennas, set $\alpha_i = \sqrt{1/N_R}$. We use MMSE criterion in both FDE and FDE-DFE schemes with equal power allocation.

*1) FDE:* The MMSE criterion for finding the FDE coefficients at the destination can be expressed as

$$\min_{\widehat{W}_{l,j}} E\left\{|\widehat{z}_m - \widehat{s}_m|^2\right\} =$$

$$\min_{\widehat{W}_{l,j}} E\left\{\left|\frac{1}{M}\sum_{l=0}^{M-1}\sum_{j=1}^{N_D}\left(\sqrt{\frac{1}{N_R}}\sum_{i=1}^{N_R} G_{l,i,j}\widehat{S}_l + \widehat{V}_{l,j}\right)\right.\right.$$
$$\left.\left. \times \widehat{W}_{l,j} \exp\left(j2\pi\frac{ml}{M}\right) - \widehat{s}_m\right|^2\right\}. \quad (19)$$

Define

$$\alpha_{eq} \triangleq \sqrt{\frac{1}{N_R}} \quad (20a)$$

and $\quad \widetilde{\mathbf{G}}_l \triangleq \left(\sum_{i=1}^{N_R} G_{l,i,1}, \sum_{i=1}^{N_R} G_{l,i,2}, \ldots, \sum_{i=1}^{N_R} G_{l,i,N_D}\right)^H,$
$$(20b)$$

where $l = 0, 1, \ldots, (M-1)$. Similar to (13), FDE coefficients can be obtained as

$$\widehat{\mathbf{W}}_l = \frac{\alpha_{eq}\widetilde{\mathbf{G}}_l}{\frac{1}{\widehat{\text{SNR}}} + \alpha_{eq}^2 \widetilde{\mathbf{G}}_l^H \widetilde{\mathbf{G}}_l}, \quad (21)$$

where $l = 0, 1, \ldots, (M-1)$ and $\widehat{\text{SNR}} = \frac{P_R}{\widehat{\sigma}_n^2}$.

*2) FDE-DFE:* In this scheme, the MMSE criterion can be expressed as

$$\min_{\widehat{W}_{l,j}} E\left\{|\widehat{z}_m - \widehat{s}_m|^2\right\} =$$

$$\min_{\widehat{W}_{l,j}} E\left\{\left|\frac{1}{M}\sum_{l=0}^{M-1}\sum_{j=1}^{N_D}\left(\alpha_{eq}\sum_{i=1}^{N_R} G_{l,i,j}\widehat{S}_l + \widehat{V}_{l,j}\right)\right.\right.$$
$$\left.\left. \times \widehat{W}_{l,j} \exp\left(j2\pi\frac{ml}{M}\right) - \sum_{k\in F_{B_g}} \widehat{f}_k^* \widehat{s}_{m-k} - \widehat{s}_m\right|^2\right\}.$$
$$(22)$$

FDE and DFE coefficients are obtained as

$$\widehat{\mathbf{W}}_l = \frac{\alpha_{eq}\widetilde{\mathbf{G}}_l}{\frac{1}{\widehat{\text{SNR}}} + \alpha_{eq}^2 \widetilde{\mathbf{G}}_l^H \widetilde{\mathbf{G}}_l}\left(1 + \sum_{k\in F_{B_g}} \widehat{f}_k^* \exp\left(-j2\pi\frac{kl}{M}\right)\right),$$
$$(23)$$

where $l = 0, 1, \ldots, (M-1)$ and

$$\widehat{\mathbf{f}} = -\widehat{\mathbf{V}}^{-1}\widehat{\mathbf{v}}, \quad (24a)$$

where

$$\widehat{\mathbf{f}} = \left(\widehat{f}_{k_1}, \widehat{f}_{k_2}, \ldots, \widehat{f}_{k_{B_g}}\right)^T, \quad (24b)$$

$$\widehat{\mathbf{v}} = \left(\widehat{v}_{k_1}, \widehat{v}_{k_2}, \ldots, \widehat{v}_{k_{B_g}}\right)^T, \quad (24c)$$

$$\widehat{\mathbf{V}} = \begin{bmatrix} \widehat{v}_0 & \widehat{v}_{k_1-k_2} & \cdots & \widehat{v}_{k_1-k_{B_g}} \\ \widehat{v}_{k_2-k_1} & \widehat{v}_0 & \cdots & \widehat{v}_{k_2-k_{B_g}} \\ \vdots & \vdots & \ddots & \vdots \\ \widehat{v}_{k_{B_g}-k_1} & \widehat{v}_{k_{B_g}-k_2} & \cdots & \widehat{v}_0 \end{bmatrix}, \quad (24d)$$

and $\quad \widehat{v}_k = \frac{1}{M \times \widehat{\text{SNR}}} \sum_{l=0}^{M-1} \frac{\exp\left(-j2\pi\frac{kl}{M}\right)}{\frac{1}{\widehat{\text{SNR}}} + \alpha_{eq}^2 \widetilde{\mathbf{G}}_l^H \widetilde{\mathbf{G}}_l}. \quad (24e)$

*B. Optimum Power Allocation*

We assume $\sum_{i=1}^{N_R} |\alpha_i|^2 \leq 1$, which means total power of the relay should be constrained to $P_R$. Therefore, by adopting the MMSE criterion, we try to solve the optimization problem for FDE and FDE-DFE schemes.

*1) FDE:* The optimization problem for this scheme is

$$\min_{\alpha_i,\widehat{W}_{l,j}} \widehat{f}_0 = \min_{\alpha_i,\widehat{W}_{l,j}} E\left\{|\widehat{z}_m - \widehat{s}_m|^2\right\} =$$

$$\min_{\alpha_i,\widehat{W}_{l,j}} E\left\{\left|\frac{1}{M}\sum_{l=0}^{M-1}\sum_{j=1}^{N_D}\left(\sum_{i=1}^{N_R} \alpha_i G_{l,i,j}\widehat{S}_l + \widehat{V}_{l,j}\right)\right.\right.$$
$$\left.\left. \times \widehat{W}_{l,j} \exp\left(j2\pi\frac{ml}{M}\right) - \widehat{s}_m\right|^2\right\},$$

s.t. $\quad \widehat{f}_1 = \sum_{i=1}^{N_R} |\alpha_i|^2 - 1 \leq 0. \quad (25)$

Due to random nature of channel coefficients, the Hessian of objective function is not always positive definite. Therefore, the objective function in (25) is generally not convex. For any optimization problem with differentiable objective and constraint functions for which strong duality holds, any pair of primal and dual optimal points must satisfy Karush-Kuhn-Tucker (KKT) conditions [9]. It is clear that our objective and constraint functions are differentiable, and thus, we try to find the primal and dual optimal points that satisfy the KKT conditions, which are expressed as

$$\begin{aligned} \nabla \widehat{f}_0 + \lambda \widehat{f}_1 &= 0, \\ \widehat{f}_1 &\leq 0, \\ \lambda &\geq 0, \\ \lambda \widehat{f}_1 &= 0. \end{aligned} \quad (26)$$

By some calculations for satisfying the KKT conditions, it can be shown that there is no solution if $\lambda = 0$. As a result, if $\widehat{f}_1 = \sum_{i=1}^{N_R}|\alpha_i|^2 - 1 = 0$, we obtain the following equations.

$$\widehat{\mathbf{W}}_l = \frac{\mathbf{U}_l}{\frac{1}{\widehat{\text{SNR}}} + \mathbf{U}_l^H \mathbf{U}_l}, \quad (27)$$

$$\lambda = \frac{1}{M \times \widehat{\text{SNR}}} \sum_{l=0}^{M-1}\sum_{j=1}^{N_D} \left|\widehat{W}_{l,j}\right|^2, \quad (28)$$

$$\boldsymbol{\alpha} = \left(M\lambda \times \mathbf{I}_{N_R\times N_R} + \sum_{l=0}^{M-1} \mathbf{C}_l\mathbf{C}_l^H\right)^{-1}\sum_{l=0}^{M-1} \mathbf{C}_l, \quad (29)$$

where $\quad \boldsymbol{\alpha} = (\alpha_1, \ldots, \alpha_{N_R})^T, \quad (30a)$

$$\mathbf{U}_l = \left(\sum_{i=1}^{N_R}\alpha_i G_{l,i,1}, \ldots, \sum_{i=1}^{N_R}\alpha_i G_{l,i,N_D}\right)^H,$$
$$(30b)$$

$$\widehat{\mathbf{W}}_l = \left(\widehat{W}_{l,1}, \ldots, \widehat{W}_{l,N_D}\right)^T, \quad (30c)$$

and
$$\mathbf{C}_l = \left(\sum_{j=1}^{N_D} \widehat{W}_{l,j} G_{l,1,j}, \ldots, \sum_{j=1}^{N_D} \widehat{W}_{l,j} G_{l,N_R,j}\right)^H, \quad (30d)$$

for $l = 0, 1, \ldots, (M-1)$.
We propose an effective algorithm for finding the optimum values.

---
**Algorithm 1**

1: Initialize $\boldsymbol{\alpha}$, such that $\sum_{i=1}^{N_R} |\alpha_i|^2 = 1$.
2: Update $\widehat{\mathbf{W}}_l$ for $l = 0, 1, \ldots, (M-1)$ by (27).
3: Update $\lambda$ by (28).
4: Update $\boldsymbol{\alpha}^{new}$ by (29).
5: If $|\alpha_i^{new} - \alpha_i| < \varepsilon$ for $i = 1, 2, \ldots, N_R$, $\widehat{\mathbf{W}}_l$ and $\boldsymbol{\alpha}^{new}$ are the desired results, otherwise set $\boldsymbol{\alpha} = \boldsymbol{\alpha}^{new}$ and go to step 2.

---

Note that if $\lambda^\circ$, $\boldsymbol{\alpha}^\circ$, and $\widehat{\mathbf{W}}_l^\circ$ for $l = 0, 1, \ldots, (M-1)$ are the optimum values, then $\lambda^\circ$, $\boldsymbol{\alpha}^\circ \exp(j\phi)$, and $\widehat{\mathbf{W}}_l^\circ \exp(-j\phi)$ for $0 < \phi < 2\pi$ and $l = 0, 1, \ldots, (M-1)$ are other sets of optimum values. Due to non-convexity feature of the optimization problem, there are multiple set of answers which obtain strong duality. Every set of answers can be achieved, depending on the initial values in the algorithm 1.

*2) FDE-DFE:* Here, the optimization problem is as follows.

$$\min_{\alpha_i, \widehat{W}_{l,j}, \widehat{f}_k} \hat{f}_0 = \min_{\alpha_i, \widehat{W}_{l,j}, \widehat{f}_k} E\left\{|\widehat{z}_m - \widehat{s}_m|^2\right\} =$$

$$\min_{\alpha_i, \widehat{W}_{l,j}, \widehat{f}_k} E\left\{\left|\frac{1}{M}\sum_{l=0}^{M-1}\sum_{j=1}^{N_D}\left(\sum_{i=1}^{N_R}\alpha_i G_{l,i,j}\widehat{S}_l + \widehat{V}_{l,j}\right)\right.\right.$$
$$\left.\left.\times \widehat{W}_{l,j}\exp\left(j2\pi\frac{ml}{M}\right) - \sum_{k\in F_{B_g}}\widehat{f}_k^*\widehat{s}_{m-k} - \widehat{s}_m\right|^2\right\},$$

s.t. $\hat{f}_1 = \sum_{i=1}^{N_R} |\alpha_i|^2 - 1 \le 0.$ (31)

The same KKT conditions in (26) must be satisfied here. Therefore, the following equations can be obtained.

$$\widehat{\mathbf{W}}_l = \frac{\mathbf{U}_l}{\frac{1}{\widehat{\text{SNR}}} + \mathbf{U}_l^H \mathbf{U}_l}\left(1 + \sum_{k\in F_{B_g}} \widehat{f}_k^* \exp\left(-j2\pi\frac{kl}{M}\right)\right), \quad (32)$$

for $l = 0, 1, \ldots, (M-1)$,

$$\lambda = \frac{1}{M \times \widehat{\text{SNR}}}\sum_{l=0}^{M-1}\sum_{j=1}^{N_D}|\widehat{W}_{l,j}|^2, \quad (33)$$

$$\widehat{f}_k = \frac{1}{M}\sum_{l=0}^{M-1}\sum_{i=1}^{N_R}\sum_{j=1}^{N_D}\alpha_i^* G_{l,i,j}^* \widehat{W}_{l,j}^* \exp\left(-j2\pi\frac{kl}{M}\right), \quad (34)$$

for $k = k_1, k_2, \ldots, k_{B_g}$,

$$\boldsymbol{\alpha} = \left(M\lambda \times \mathbf{I}_{N_R \times N_R} + \sum_{l=0}^{M-1} \mathbf{C}_l \mathbf{C}_l^H\right)^{-1} \sum_{l=0}^{M-1} \mathbf{C}_l$$
$$\times \left(1 + \sum_{k\in F_{B_g}} \widehat{f}_k^* \exp\left(-j2\pi\frac{kl}{M}\right)\right). \quad (35)$$

An effective algorithm for finding the optimum values is presented as follows.

---
**Algorithm 2**

1: Initialize $\boldsymbol{\alpha}$, such that $\sum_{i=1}^{N_R} |\alpha_i|^2 = 1$.
2: Initialize $\widehat{\mathbf{f}}$ with zeros.
3: Update $\widehat{\mathbf{W}}_l$ for $l = 0, 1, \ldots, (M-1)$ by (32).
4: Update $\lambda$ by (33).
5: Update $\widehat{f}_k^{new}$ for $k = k_1, k_2, \ldots, k_{B_g}$ by (34).
6: Update $\boldsymbol{\alpha}^{new}$ by (35).
7: If $|\alpha_i^{new} - \alpha_i| < \varepsilon$ for $i = 1, 2, \ldots, N_R$ and $|\widehat{f}_k^{new} - \widehat{f}_k| < \varepsilon$ for $k = k_1, k_2, \ldots, k_{B_g}$, then $\widehat{\mathbf{W}}_l$ for $l = 0, 1, \ldots, (M-1)$, $\widehat{\mathbf{f}}^{new}$, and $\boldsymbol{\alpha}^{new}$ are the desired results, otherwise set $\boldsymbol{\alpha} = \boldsymbol{\alpha}^{new}$, $\widehat{\mathbf{f}} = \widehat{\mathbf{f}}^{new}$, and go to step 3.

---

## V. SIMULATION RESULTS

In this section, some numerical results are provided. Throughout our simulations, we assume quaternary phase shift keying (QPSK) modulation and equal noise variances at the relay and destination ($\sigma_n^2 = \widehat{\sigma}_n^2$). Also, we suppose $M = 512$ symbols in every transmission and use SNR as we defined in (12a). Moreover, the channel power delay profile is assumed to be

$$p[n] = \frac{\overline{P}}{\sigma_t}\sum_{l=0}^{L_x - 1} e^{-\frac{n}{\sigma_t}}\delta[n - lT_s], \quad (36)$$

where $\overline{P}$ represents the average power of the multipath components, $\sigma_t$ indicates the delay spread, $L_x \in \{L_h, L_g\}$, $T_s$ represents the symbol duration, and the channel taps are zero-mean complex Gaussian random variables. We assume $P_S = P_R, \overline{P} = 1, T_s = 1, L_{CP} = 20, B_h = L_h - 1, B_g = L_g - 1$, and $\varepsilon = 0.001$ in our simulations. For convenience, DFE is considered at both the relay and destination or in none of them.

In Fig. 2, the effect of the number of antennas at the relay and destination on the bit error rate (BER) is investigated for SC-FDE with equal power allocation. We choose the delay spread factor and the length of channels as $\sigma_t = 2$ and $L_h = L_g = 3$, respectively. As can be observed, increasing $N_R$ and $N_D$, improves the BER performance. If $2N$ antennas are considered at both the relay and destination, picking $N$ antennas for relay and destination is the best choice in achieving a superior BER performance. Hence, in the rest of our simulations, we use equal number of antennas at the relay and destination.

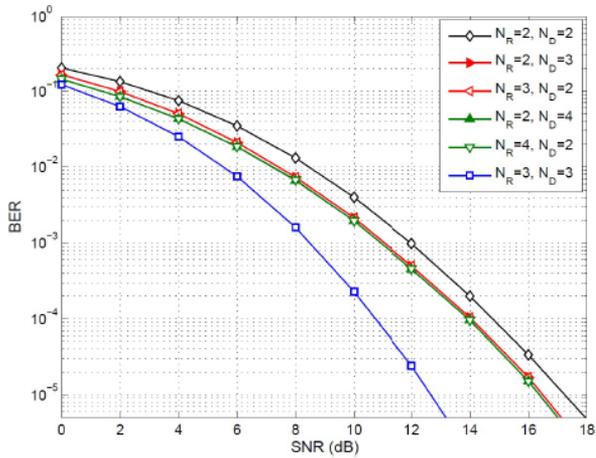

Fig. 2. BERs of SC-FDE with equal power allocation for various number of relay and destination antennas. $\sigma_t = 2$ and $L_h = L_g = 3$.

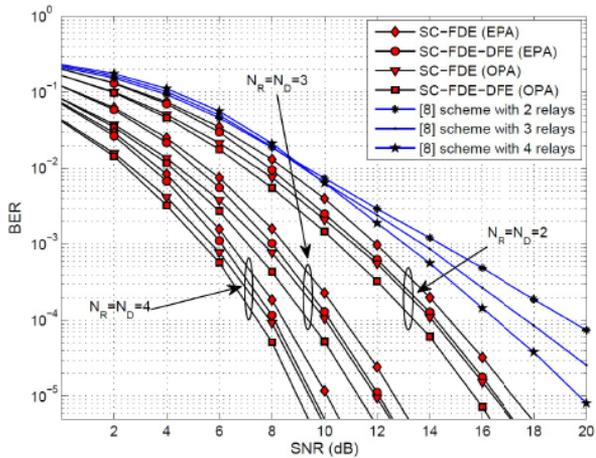

Fig. 3. BERs of SC-FDE and SC-FDE-DFE with equal and optimum power allocations for several number of relay and destination antennas. $\sigma_t = 2$ and $L_h = L_g = 3$. For comparison, the BERs of the proposed scheme in [8] for several number of relays are also shown.

The BERs for SC-FDE and SC-FDE-DFE with equal and optimum power allocations for various number of antennas at the relay and destination are depicted in Fig. 3. Also, the BER of the scheme in [8] has been plotted. The delay spread factor and the length of channels are the same as in Fig. 2. It is illustrated that optimum power allocation results in better performance compared to equal power allocation, and SC-FDE-DFE is better than SC-FDE in BER performance. Comparing with the scheme in [8], significant gains are obtained in our schemes. In Fig. 3 and Fig. 4, EPA and OPA denote equal and optimum power allocations, respectively.

The effect of $\sigma_t$ on the BER performance of SC-FDE and SC-FDE-DFE with equal and optimum power allocations is investigated in Fig. 4. The number of antennas at the relay and destination, the length of channels, and the SNR are chosen as $N_R = N_D = 3$, $L_h = L_g = 21$, and SNR $= 10$ dB, respectively. The results show that for our schemes, increasing $\sigma_t$, leads to lower BER. In contrast, the performance of the proposed scheme in [8] is almost invariant to the value of $\sigma_t$.

In Fig. 3 and Fig. 4, the total power of the relay in our schemes is equal to the power of each relay in the proposed scheme in [8]. Hence, higher gain will be obtained, if the total transmit power at the relays are equal in [8] and our schemes.

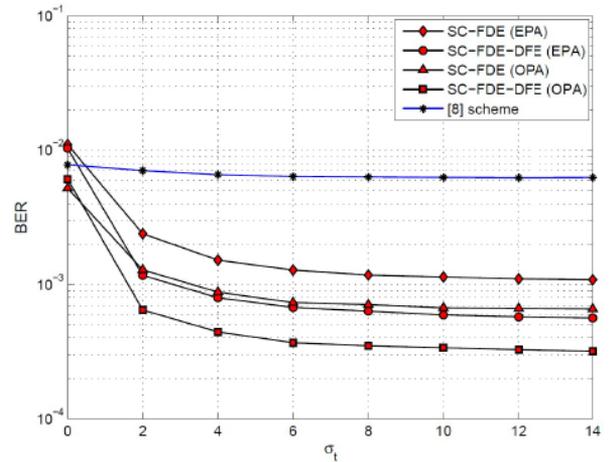

Fig. 4. BERs of SC-FDE and SC-FDE-DFE with equal and optimum power allocations vs. delay spread factor $\sigma_t$. $N_R = N_D = 3$, $L_h = L_g = 21$, and SNR $= 10$ dB. For comparison, the BER of the proposed scheme in [8] is also shown.

## VI. CONCLUSION

In this paper, we considered DF relay networks with one source, one multi-antenna relay, and multi-antenna destination in frequency-selective channels. We employed SC-FDE with or without decision feedback filter to combat the ISI in highly time dispersive channels. The relaying schemes were under equal and optimum power allocation assumptions for constant total transmit power at the relay. Using MMSE criterion, optimum filter taps and assigned coefficients for power allocation were calculated. Without requiring full CSI at the relay and destination, it is illustrated that in DF relay networks which use SC-FDE, our relaying scheme has superior performance compared to previous schemes.